# Hard X-ray Transient Grating Spectroscopy on Bismuth Germanate

Jérémy R. Rouxel[1,2,3,*], Danny Fainozzi[4], Roman Mankowsky[2], Benedikt Rösner[2], Gediminas Seniutinas[2], Riccardo Mincigrucci[4], Sara Catalini[5], Laura Foglia[4], Riccardo Cucini[6], Florian Döring[2], Adam Kubec[2], Frieder Koch[2], Filippo Bencivenga[4], Andre Al Haddad[2], Alessandro Gessini[4], Alexei A. Maznev[7], Claudio Cirelli[2], Simon Gerber[2], Bill Pedrini[2], Giulia F. Mancini[1,2,8], Elia Razzoli[2], Max Burian[2], Hiroki Ueda[2], Georgios Pamfilidis[2], Eugenio Ferrari[2], Yunpei Deng[2], Aldo Mozzanica[2], Philip J. M. Johnson[2], Dmitry Ozerov[2], Maria Grazia Izzo[9,10], Cettina Bottari[4], Christopher Arrell[2], Edwin James Divall[2], Serhane Zerdane[2], Mathias Sander[2], Gregor Knopp[2], Paul Beaud[2], Henrik T. Lemke[2], Christopher J. Milne[2,11], Christian David[2], Renato Torre[5], Majed Chergui[1], Keith A. Nelson[7], Claudio Masciovecchio[4], Urs Staub[2], Luc Patthey[2] and Cristian Svetina[2,*]

1 Ecole Polytechnique Fédérale de Lausanne, Laboratory of Ultrafast Spectroscopy (LSU) and Lausanne Centre for Ultrafast Science (LACUS), CH-1015 Lausanne, Switzerland
2 Paul Scherrer Institut, Forschungsstrasse 111, 5232 Villigen, Switzerland
3 Univ Lyon, UJM-Saint-Etienne, CNRS, Graduate School Optics Institute, Laboratoire Hubert Curien UMR 5516, Saint-Etienne F-42023, France
4 Elettra-Sincrotrone Trieste S.C.p.A., S.S. 14 km 163,5 in Area Science Park, I-34149 Basovizza, Trieste, Italy
5 European Laboratory for Non-Linear Spectroscopy (LENS) and Dip. Di Fisica ed Astronomia, Università degli studi di Firenze, 50019, Sesto Fiorentino, Firenze, Italy
6 Istituto Officina dei Materiali-CNR, 34149 Basovizza (TS), Italy
7 Department of Chemistry, Massachusetts Institute of Technology, Cambridge, Massachusetts 02139, USA
8 Department of Physics, University of Pavia, I-27100 Pavia, Italy
9 Sapienza Università di Roma, Dipartimento di Ingegneria informatica, automatica e gestionale Antonio Ruberti, via Ariosto 25 00185 Roma, Italy
10 Istituto Italiano di Tecnologia-Center for Life Nanoscience, Viale Regina Elena, 291, 00161 Roma, Italy
11 European XFEL GmbH, Holzkoppel 4, 22869 Schenefeld, Germany

E-mail: cristian.svetina@psi.ch; jeremy.rouxel@univ-st-etienne.fr;

**Optical-domain Transient Grating (TG) spectroscopy is a versatile background-free four-wave-mixing technique used to probe vibrational, magnetic and electronic degrees of freedom in the time domain[1]. The newly developed coherent X-ray Free Electron Laser sources allow its extension to the X-ray regime. X-rays offer multiple advantages for TG: their large penetration depth allows probing the bulk properties of materials, their element-specificity can address core-excited states, and their short wavelengths create excitation gratings with unprecedented momentum transfer and spatial resolution. We demonstrate for the first time TG excitation in the hard X-ray range at 7.1 keV. In Bismuth Germanate (BGO), the non-resonant TG excitation generates coherent optical phonons detected as a function of time by diffraction of an optical probe pulse. This experiment demonstrates the ability to probe bulk properties of materials and paves the way for ultrafast coherent four-wave-mixing techniques using X-ray probes and involving nanoscale TG spatial periods.**

Four-wave mixing (FWM), in which three coherent pulsed beams with controlled wavelengths, polarizations, wavevectors and arrival times are crossed at the sample[2,3], is among the most used optical configurations in time-resolved non-linear spectroscopies. The non-linear response of the sample to the excitation induced by the crossed light fields generates a fourth beam that emerges with a wavevector determined by the incident wavevectors. A specific outcome of non-linear spectroscopies is two-dimensional spectroscopies[4], which can measure directly matter coherences and disentangle homogeneous and inhomogeneous contributions to spectral lineshapes.

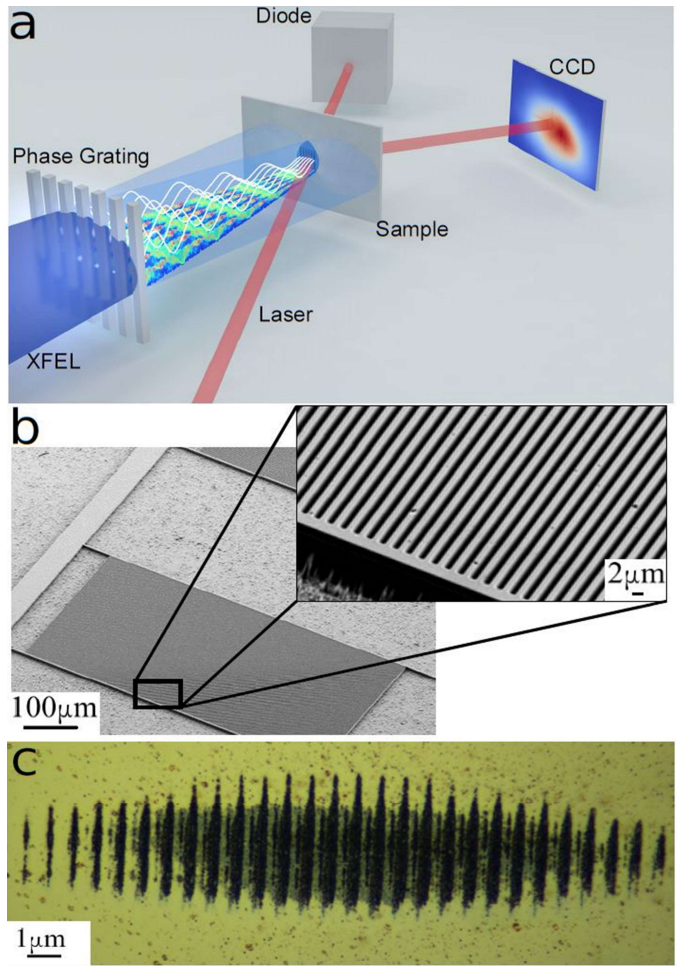

*Figure 1: a) Schematic of the XTG setup. The incoming X-ray FEL pulse (in blue) is diffracted by a transmission phase grating. Interferences between the diffracted orders generate a Talbot carpet, i.e. a region of self-images of the originating phase grating. The sample experiences a TG excitation with smaller spatial periodicity due to the convergence of the X-ray beam. A delayed optical pulse (in red) probes the dynamics stimulated by the X-ray TG via transient diffraction, detected by an array detector (CCD) b) Scanning Electron Microscope (SEM) image of the diamond phase grating. c) Permanent grating imprint on a gold surface placed at the sample position and illuminated with an intense XFEL pulse.*

A widely used optical configuration in time-resolved non-linear spectroscopy is the transient grating geometry (see Fig.1a) in which two simultaneous pulses are crossed in the sample to generate an optical interference pattern that leads to a grating of spatially modulated excitation. A controlled time-delayed probe pulse is then diffracted by this grating. The time-dependence of the signal reveals the sample dynamics. The spatial period of the interference pattern is $\Lambda=\lambda/2\sin\theta$, with $\lambda$ the pump wavelength and $2\theta$ the relative angle between the two pulses. In reciprocal space, the TG wavevector **q**, with magnitude $q = 2\pi/\Lambda$, is the difference between the two excitation wavevectors **k₁** and **k₂**. The diffracted signal wavevector **k_s** is given by the phase-matching condition **k_s**=**k₁**-**k₂**+**k₃** where **k₃** is the probe beam wavevector. TGs have been used to investigate dispersive excitations such as acoustic waves[5], phonon-polaritons[6], transport phenomena in solids, including heat, charge, chemical species, excitons, and spins[7,8,9,10]. These measurements are often conducted in a range of $\Lambda$ matching the frequency of acoustic waves in the sample[11]. TG can be further used to distinguish between diffusive or ballistic transport as it allows for the discrimination between long and short mean free path scattering events in the quasiparticle transport. For example, thermal transport can be readily measured in various materials by TG following the temporal evolution of the signal. A TG density modulation in a bulk material produces a refractive index change that diffracts the probe light. The diffracted signal intensity diminishes in time as heat moves from the TG peaks to minima, ultimately suppressing the TG pattern and hence the scattered signal. Notice that in the case of an opaque material for the excitation frequency, a TG modulation corresponds to a surface modulation rather than a bulk modulation.

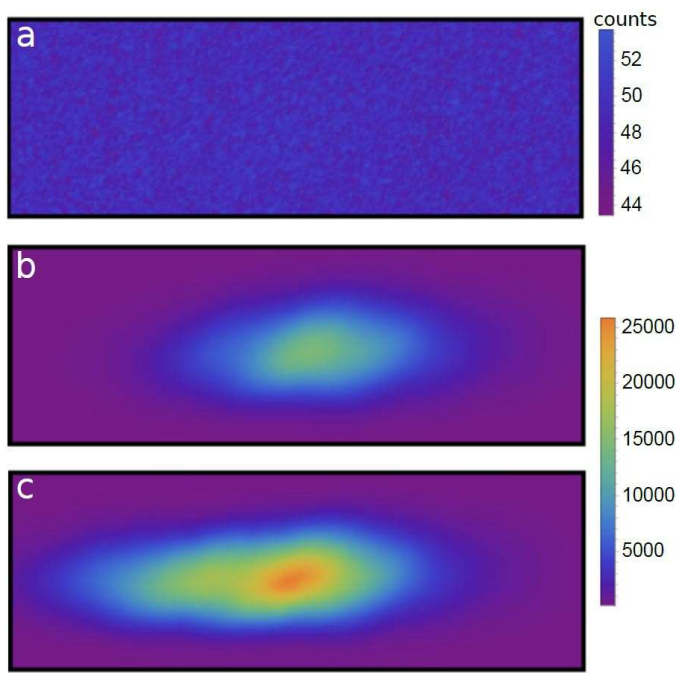

*Figure 2: Footprints of the signal beam on the CCD detector. Before time-zero (a), no signal is observed. At time zero (b), the XTG signal rises and reaches a maximum after 150 fs (c).*

In high-quality crystals, thermal transport was observed to be diffusive for TG periods of tens of microns, with ballistic contributions leading to deviations from diffusive kinetics at TG periods below 15 μm, i.e. comparable to the mean free path of some of the heat-carrying acoustic phonons[12,13]. At such a spatial range, the TG signal decay timescale $\tau$ due to transport was proportional to the square of the spatial period, which determines the transport length $D$, i.e. $D \propto \sqrt{\tau}$. In most materials, submicron spatial periods are necessary to see non-diffusive thermal kinetics, and nanoscale periods are required in order to resolve the ballistic limit. New information and insights about other transport properties can also be provided by measurements with nanometric TG periods. GHz and THz acoustic modes can be observed, even if they are highly damped as is common at high frequencies in imperfect or disordered materials. TG periods matching or approaching those of charge or spin-density waves or structural modulations might result in excitation of amplitude or phase modes. There is thus great incentive to develop nanoscale TG methodology which requires short wavelengths. The availability of X-ray free-electron lasers (XFELs) makes it possible to generate gratings down to few nanometers periods. In the hard X-ray region, TG measurements could explore material wavevector ranges currently reached in inelastic X-ray scattering and inelastic neutron scattering (IXS and INS)[14]. We believe that this will add a complementary tool in many research fields, just as optical TG and FWM measurements have provided a wealth of information that could not be extracted from light scattering methods in the same spectral range.

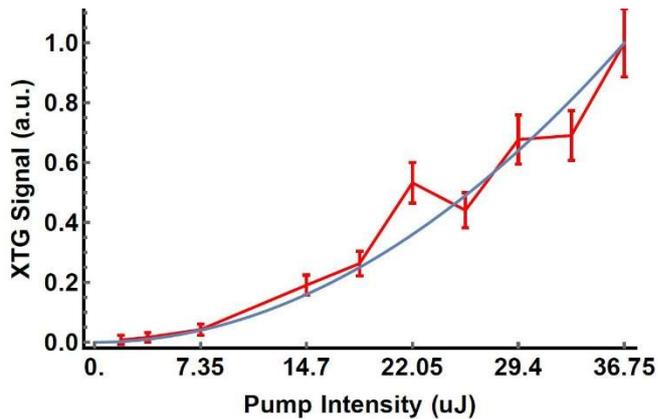

*Figure 3: XTG signal at 2 ps time delay as a function of the X-ray intensity at the sample. The experimental results (in red) are consistent with a quadratic trend (in blue).*

Recently, TG in the extreme ultraviolet (EUV) was successfully demonstrated[15,16]. However, EUV light penetrates matter at most a few tens of nanometers and restricts resonant excitation to light elements. Extending TG into the hard X-ray range (XTG) would allow probing bulk material behavior[17], further increasing the range of TG wavevectors that could be reached, and giving access to the resonant excitation at the absorption edges of a broader range of elements.

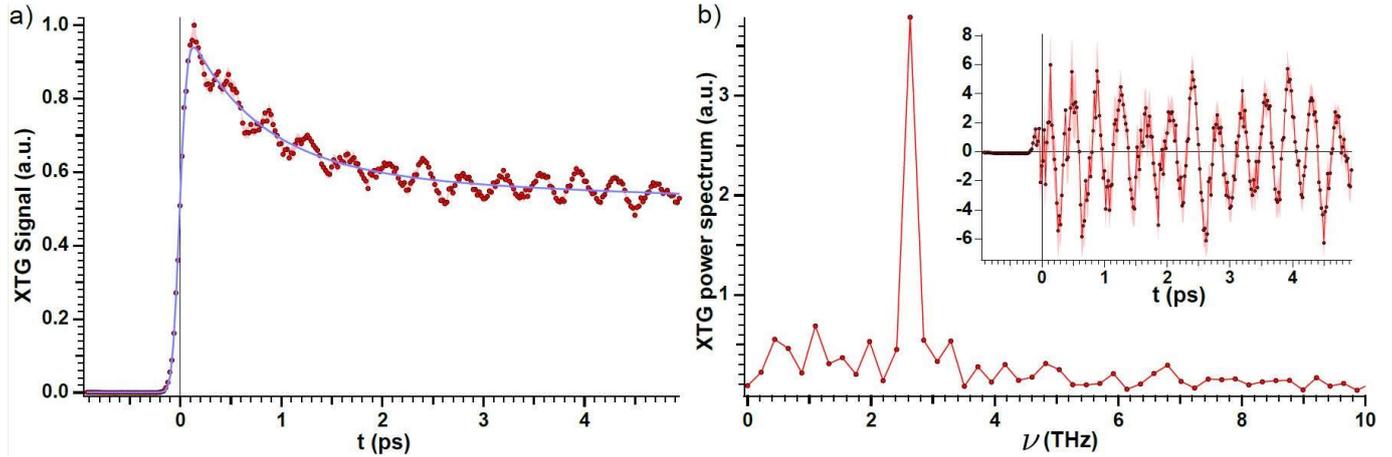

*Figure 4: XTG signals from BGO at 7.1 keV with an excitation grating pitch of 770 nm. a) Time trace at short times (<5 ps) normalized to 1. The non-oscillatory decay of the signal is fitted (in blue) with Eq. 2. The fast oscillations are attributed to optical phonons coherence. b) Fourier transform of the fast-oscillatory component (residuals from the fit in a) in inset).*

Crossing the beams to generate and probe a TG becomes very challenging in the hard X-ray range. Having recently demonstrated the ability to use the Talbot effect to produce permanent spatial gratings in gold, we demonstrate here its use in the weak interaction limit in BGO. This offers an optimal approach to investigate ultrafast transport with very high line-density XTG, while avoiding the use of a grazing incidence geometry on reflective optics responsible for substantial losses and complicated geometries. The experimental layout at the SwissFEL[19] Bernina endstation[20] is shown in Figure 1a. It employs a diamond phase grating with a period of 960 nm in the X-ray beam path (Fig.1b). Diamond is used for its resilience to high intensity X-ray pulses. The diffraction on the grating generates the spatially and temporally overlapped phase-locked interfering beams avoiding the complex procedure of splitting and crossing X-ray beams. A variably delayed optical probe pulse is incident at the Bragg angle for diffraction from the TG material response. The XTG was created using 7.1 keV, 40 fs (rms) duration pulses from the SwissFEL[19,20] without a monochromator. The defocused X-ray beam produced an elliptical 250 μm x 150 μm (FWHM) spot at the sample, a room-temperature crystal of bismuth germanium oxide, $Bi_4Ge_3O_{12}$ (BGO). The diamond phase grating was placed 150 mm upstream of the sample to generate the XTG excitation gratings of 770 nm period, as seen on a gold target using intensities above the damage threshold (Fig.1c). This perfectly agrees with the simulated Talbot carpet pattern. A 400-nm, 80 fs (rms) laser pulse (3.1 eV, second harmonic of a Ti:Sapphire laser) was delivered onto the sample with a spot size of 190 μm x 150 μm (FWHM). The energies of the FEL and probe laser pulses at the sample were 1.5 μJ and 1.2 µJ, respectively. The diffracted probe beam was recorded by a visible CCD camera mounted on a diffractometer. The signal was obtained by integrating the diffracted light intensity at each probe delay over a detector positioned at the diffraction (phase matching) angle. Fig.2 shows typical camera images detected before (a), at (b), and 150 fs after (c) the overlap time between the XTG and the probe (time zero). To reduce background, a lens was installed in front of the CCD and the observed spot shape results from both our imaging setup and the spread due to the detector angle of the sample. Further details on the experimental procedures are given in the Supplementary Materials.

BGO is an optically isotropic material with cubic crystalline structure (eulitine), which is widely used for its numerous applications[21,22], such as scintillation detectors for high-energy physics[23,24], holographic data storage material[24], high-resolution positron emission tomography[25], solid-state laser host when activated with trivalent rare-earth ions[26,27], and as an electro-optic material for optical voltage, current, and electric power sensors[28].

Fig.3 shows the dependence of the signal intensity $I_{XTG}$ on the X-ray pump intensity $I_{FEL}$ at a 2 ps probe time delay. The quadratic dependence confirms the expected signal nonlinearity[5]:

$$I_{XTG} \propto \left| \chi^{(3)} I_{FEL} I_{probe}^{1/2} \frac{L}{\lambda} \text{sinc}(\Delta k\, L/2) \right|^2 \quad (1)$$

where $\chi^{(3)}$ is the effective third-order susceptibility, $I_{FEL}$ and $I_{probe}$ are the FEL pump and probe light intensity respectively, $L$ is the interaction length in the sample, $\lambda$ is the probe wavelength and $\Delta k$ is the phase matching condition,. Our diffraction efficiency is considerably higher than in the EUV (see SM §5.5 for further details) because of the greater penetration depth (approximately 6 µm) of hard X-rays (BGO is transparent to the near-UV probe wavelength), highlighting the capability of XTG to measure bulk properties of matter.

In Fig.4a, the XTG time trace displays a sharp rise at time zero ($t_0$) followed by oscillations and a decay to a steady-state signal. The non-oscillatory part of the signal can be fit by the following time-dependent form via least-square minimization:

$$I(t) = \left| \frac{1}{2}\left(1 + \text{erf}\left(\frac{t-t_0}{\sigma}\right)\right)\left(c_1 e^{-\frac{t-t_0}{\tau}} + c_2\right) \right|^2 \quad (2)$$

where $c_1$ is the amplitude of the electronic to A1 optical phonon mode energy transfer that decays with t = 970 ± 30 fs and $c_2$ accounts for a thermal signal that is constant on the displayed timescale. The quantity σ = 92 ± 6 fs is the rise time, reflecting the convolution between the X-ray pulse duration and the probe pulse duration as well as the non-collinear geometry. The periodic oscillation (emphasized in the inset of Fig.4b) corresponds to the $A_1$ optical phonon mode of BGO at a frequency of 2.6 ± 0.1 THz in very good agreement with the literature[29,30] and with the phonon dispersion curve at the XTG wavevector (see SM § 3). The optical phonon and the single exponential decay of the steady-state signal have also been reported in the EUV experiment[31]. Data extending to 300 ps delay are shown in the SM and possess 24 GHz oscillations. These oscillations, which are assumed to have a thermal origin, require further investigations. XTG signals are able to follow the material response from tens of femtoseconds to tens of nanoseconds, thus spanning a broad range of physical phenomena.

The relatively simple XTG approach used here can easily be implemented at any XFEL facility on solids or liquids, and to probe a broad range of materials such as magnetic systems, heterostructures, and excitonic systems[32] such as light harvesting complexes, to name a few. Indeed, the use of hard X-ray TG offers various advantages compared to optical or EUV TG. First, extremely small excitation grating periods (high TG wavevectors) are now achievable. In the present case, the main limitation is the wavelength of the optical probe, which puts a lower bound on the periodicity of the TG that can be probed. However, a hard X-ray probe can easily detect periods at the nanometric scales or even down to the ångström. Further limitations could arise from the phase grating pattern. Here, nanolithography techniques for the fabrication of diffractive hard X-ray lenses with periods of tens of nanometers have been developed[33,34]. It may also become possible to use superlattice structures for the phase grating[35] in order to move to even smaller periods, or to design other means of crossing the X-ray beams such that large angles θ can be achieved. The present phase grating period range can already provide access to bulk nanoscale heat, charge, and spin transport that are difficult to access through conventional methods. The capability for rapidly changing XTG wavevector by switching the phase grating patterns, as routinely done in optical TG experiments, will enable incisive transport measurements in the nanoscale range where macroscale diffusive kinetics may no longer apply[36,37]. Elemental specificity will be extremely useful for electronic and spin measurements. With the ultrashort pulses of new X-ray sources (few-fs or below), measurements of the coherent relaxation of core-hole states can become accessible, adding valuable information on the core excited state dynamics. The signal sensitivity could also be largely improved by using self-heterodyne detection schemes that make use of the probe diffracted beam from the phase grating as a local oscillator[38,39]. Finally, the Talbot approach can be readily extended to multi-color FEL pulses for different excitation and probe wavelengths. This will facilitate the implementation of condensed phase analogues of theoretically proposed methods[40] for the study of the delocalization of electronic excited states, electronic coherences and charge transfer processes, with ultrafast time resolution and atomic selectivity.

**References**

1 Goodno, G. D., Dadusc, G., & Miller, R. D. (1998). Ultrafast heterodyne-detected transient-grating spectroscopy using diffractive optics. *JOSA B*, *15*(6), 1791-1794.

2 Thomson, R., C. Leburn, and D. Reid, eds. *Ultrafast nonlinear optics*. Springer, (2013)

3 Mukamel, S. *Principles of nonlinear optical spectroscopy*. Vol. 6. New York: Oxford university press, (1995)

4 Hamm, Peter, and Martin Zanni. *Concepts and methods of 2D infrared spectroscopy*. Cambridge University Press, (2011)

5 Rogers, J. A., Maznev, A. A., Banet, M. J., & Nelson, K. A. (2000). Optical generation and characterization of acoustic waves in thin films: Fundamentals and applications. *Annual Review of Materials Science*, *30*(1), 117-157.

6 Crimmins, T. F., Stoyanov, N. S., & Nelson, K. A. (2002). Heterodyned impulsive stimulated Raman scattering of phonon–polaritons in LiTaO 3 and LiNbO 3. *The Journal of chemical physics*, *117*(6), 2882-2896.

7 Redman, D. A., et al. "Spin dynamics and electronic states of N-V centers in diamond by EPR and four-wave-mixing spectroscopy." *Physical review letters* 67.24 (1991): 3420

8 West, B. A., J. M. Womick, and A. M. Moran. "Probing ultrafast dynamics in adenine with mid-UV four-wave mixing spectroscopies." *The Journal of Physical Chemistry A* 115, no. 31 (2011): 8630-8637

9 Dhar, L., J. A. Rogers, and K. A. Nelson. "Time-resolved vibrational spectroscopy in the impulsive limit." *Chemical Reviews* 94, no. 1 (1994): 157-193

10 Janušonis, J., T. Jansma, C. L. Chang, Q. Liu, A. Gatilova, A. M. Lomonosov, V. Shalagatskyi, T. Pezeril, V. V. Temnov, and R. I. Tobey. "Transient grating spectroscopy in magnetic thin films: Simultaneous detection of elastic and magnetic dynamics." *Scientific reports* 6 (2016): 29143

11 Tobey, R. I., Siemens, M. E., Murnane, M. M., Kapteyn, H. C., Torchinsky, D. H., & Nelson, K. A. (2006). Transient grating measurement of surface acoustic waves in thin metal films with extreme ultraviolet radiation. *Applied physics letters*, *89*(9), 091108.

12 Johnson, J. A., Maznev, A. A., Cuffe, J., Eliason, J. K., Minnich, A. J., Kehoe, T., ... & Nelson, K. A. (2013). Direct measurement of room-temperature nondiffusive thermal transport over micron distances in a silicon membrane. *Physical review letters*, *110*(2), 025901.

13 Hua, C., and Austin J. Minnich. "Transport regimes in quasiballistic heat conduction." *Physical Review B* 89, no. 9 (2014): 094302

14 Behrens, C., Decker, F. J., Ding, Y., Dolgashev, V. A., Frisch, J., Huang, Z., ... & Turner, J. (2014). Few-femtosecond time-resolved measurements of X-ray free-electron lasers. *Nature Communications*, *5*(1), 1-7.

15 Bencivenga, F., Cucini, R., Capotondi, F. *et al.* Four-wave mixing experiments with extreme ultraviolet transient gratings. *Nature* **520,** 205–208 (2015)

16 Bencivenga, Filippo*, et al*. Nanoscale transient gratings excited and probed by extreme ultraviolet femtosecond pulses." *Science advances* 5. eaaw5805 7 (2019)

17 Schweigert, I. V., and S. Mukamel. "Coherent ultrafast core-hole correlation spectroscopy: X-ray analogues of multidimensional NMR." *Physical review letters* 99, no. 16 (2007): 163001

18 Svetina, C., R. Mankowsky, G. Knopp, F. Koch, G. Seniutinas, B. Rösner, A. Kubec et al. "Towards X-ray transient grating spectroscopy." *Optics letters* 44, no. 3 (2019): 574-577

19 Milne, C. J., T. Schietinger, M. Aiba, A. Alarcon, J. Alex, A. Anghel, V. Arsov et al. "SwissFEL: the Swiss X-ray free electron laser." *Applied Sciences* 7, no. 7 (2017): 720

20 Ingold, G., R. Abela, C. Arrell, P. Beaud, P. Böhler, M. Cammarata, Y. Deng et al. "Experimental station Bernina at SwissFEL: condensed matter physics on femtosecond time scales investigated by X-ray diffraction and spectroscopic methods." *Journal of synchrotron radiation* 26, no. 3 (2019)

21 Raymond, S.G.; Townsend, P.D. The influence of rare- earth ions on the low-temperature thermoluminescence of $Bi_4Ge_3O_{12}$, *J. Phys. Cond. Mat.,* v. 12, p. 2103-122, 2000

22 Williams, P.A.; Rose, A.H.; Lee, K.S.; Conrad, D.C.; Day, G.W.; Hale, P.D. Optical, thermo-optic, electro-optic, and photoelastic properties of bismuth germanate ($Bi_4Ge_3O_{12}$), *Appl. Opt.,* v. 35, n. 19, p. 3562-69, 1996

24 Kaminskii, A. A., Sarkisov, S. E., Denisenko, G. A., Ryabchenkov, V. V., Lomonov, V. A., Perlin, Y. E., ... & Reiche, P. (1984). Growth, spectral and luminescence study of cubic Bi4Ge3O12: Pr3+ crystals. *physica status solidi (a)*, *85*(2), 553-567.

24 Brunner, S. E., and D. R. Schaart. "BGO as a hybrid scintillator/Cherenkov radiator for cost-effective time-of-flight PET." *Physics in Medicine & Biology* 62, no. 11 (2017): 4421

25 Tao, L., Coffee, R. N., Jeong, D., & Levin, C. S. (2021). Ionizing photon interactions modulate the optical properties of crystals with femtosecond scale temporal resolution. *Physics in Medicine & Biology*, *66*(4), 045032.

26 Kamada, O., & Kakishita, K. (1993). Electro-optical effect of Bi4Ge3O12 crystals for optical voltage sensors. *Japanese journal of applied physics*, *32*(9S), 4288.

27 Kaminskii, A.A., Sarkisov, S.E., Butaeva, T.I., Denisenko, G.A., Hermoneit, B., Bohm, J., Grosskreutz, W. and Schultze, D. (1979), Growth, spectroscopy, and stimulated emission of cubic $Bi_4Ge_3O_{12}$ crystals doped with $Dy^{3+}$, $Ho^{3+}$, $Er^{3+}$, $Tm^{3+}$, or $Yb^{3+}$ ions. phys. stat. sol. (a), 56: 725-736

28 Li, C., & Yoshino, T. (2002). Simultaneous measurement of current and voltage by use of one bismuth germanate crystal. *Applied optics*, *41*(25), 5391-5397.

29 Chen, Z., Y. Gao, B. C. Minch, and M. F. DeCamp. "Coherent optical phonon generation in Bi3Ge4O12." *Journal of Physics: Condensed Matter* 23, no. 38 (2011): 385402

30 Couzi, M., J. R. Vignalou, and G. Boulon. "Infrared and Raman study of the optical phonons in Bi4Ge3O12 single crystal." *Solid State Communications* 20, no. 5 (1976): 461-465

31 Maznev, A. A., F. Bencivenga, Andrea Cannizzo, F. Capotondi, R. Cucini, R. A. Duncan, Thomas Feurer, et al. "Generation of coherent phonons by coherent extreme ultraviolet radiation in a transient grating experiment." *Applied physics letters* 113, no. 22 (2018): 221905

32 Norman, Patrick, and Andreas Dreuw. "Simulating x-ray spectroscopies and calculating core-excited states of molecules." *Chemical reviews* 118, no. 15 (2018): 7208-7248 33 Miao, H., Gomella, A. A., Chedid, N., Chen, L., & Wen, H. (2014). Fabrication of 200 nm period hard X-ray phase gratings. *Nano letters*, *14*(6), 3453-3458.

34 Vila-Comamala, Joan, et al. "Ultra-high resolution zone-doubled diffractive X-ray optics for the multi-keV regime." *Optics express* 19.1 (2011): 175-184.

35 Lynch, S. K., et al. "Fabrication of 200 nm period centimeter area hard x-ray absorption gratings by multilayer deposition." *Journal of Micromechanics and Microengineering* 22.10 (2012): 105007.

36 Cho, J., Hwang, T. Y., & Zewail, A. H. (2014). Visualization of carrier dynamics in p (n)-type GaAs by scanning ultrafast electron microscopy. *Proceedings of the National Academy of Sciences*, *111*(6), 2094-2099.

37 Gorfien, M., et al. (2020). Nanoscale thermal transport across an GaAs/AlGaAs heterostructure interface. *Structural Dynamics*, *7*(2), 025101.

38 Grilj, J., et al. (2015). Self-referencing heterodyne transient grating spectroscopy with short wavelength. *Photonics* 2 (2), 392-401).

39 Katayama, K., Yamaguchi, M., & Sawada, T. (2003). Lens-free heterodyne detection for transient grating experiments. *Applied physics letters*, *82*(17), 2775-2777.

40 Tanaka, S., and Mukamel, S. (2002). Coherent X-ray Raman spectroscopy: a nonlinear local probe for electronic excitations. *Physical review letters*, *89*(4), 043001.

**Acknowledgements** This study was supported by the the Swiss National Science Foundation (SNSF), grant No. 200021_165550/1, by the SNSF research instrument NCCR Molecular Ultrafast Science and Technology (NCCR MUST), grants No. 51NF40-183615 and No. 200021_169017, by the ERC Grant "DYNAMOX" No. ERC-2015-AdG-694097 and by the EU-H2020 Research and Innovation Programme under the Marie Skłodowska-Curie grant agreements No. 701647, No. 654360 NFFA-Europe, No. 801459-FP-RESOMUS and No. 871124 Laserlab-Europe. The contribution of the MIT participants A. A. M. and K. A. N. was supported by the U.S. Department of Energy Award DE-SC0019126. The authors thank M. Dzambegovic for the graphical rendering of Figure 1a.

**Author contributions** CS conceptualized the framework of this experiment. JR and CS designed the experiment. GS and CD fabricated the diamond gratings. BR carried out the optical microscopy of the static printed gratings. The whole team participated to the experiment and was involved in the discussions. JR, DF, EF carried out the data reduction. JR, DF and CS performed the data analysis. CS, JR and DF wrote the manuscript.

**Competing interests** The authors declare no competing financial interests.

## Methods

**Experimental setup** The incident X-ray beam is focused with bendable KB mirrors and impinges on a phase grating. The Talbot effect creates a periodic pattern (Talbot carpet) at periodic distances (Talbot distances) by the interference of the diffracted orders. Moreover, the convergence of the X-ray beam ensures that the excitation grating has a smaller pitch than the phase grating (de-magnification factor) and this pitch is controlled by the X-ray focusing, the distance between the focus and the phase grating and relative distance between grating and sample. For π/2 phase grating, the transmitted 0th order is present; it also induces some interaction in the sample, but it does not contribute to the diffracted XTG signal due to the phase-matching condition. This effect is not present in a perfect π phase grating.

In our experiment, the gratings and the sample were mounted on the General-Purpose Station of the Bernina endstation at SwissFEL with a separation of 150 mm. For a π phase grating of 1650 nm pitch, this configuration led to an excitation grating of about 660 nm pitch at the sample position. In this case, the relative separation of the Talbot planes in the sample area was about 5 mm.

**X-ray pump-beam preparation** High intensity horizontally polarized X-ray pulses were delivered by SwissFEL with time duration of about 40 fs (rms) and a repetition rate of 50 Hz. The FEL was tuned to 7.1 keV, the emitted radiation had a bandwidth of about 0.3% and was used without a monochromator (pink beam condition) for all the BGO measurements. The FEL beam was focused on a scintillator coupled with a CCD camera through a microscope (X-ray eye) 750 mm downstream the grating by tuning the curvature of the focusing Kirkpatrick-Baez (KB) mirrors. Then, the vertical focus was adjusted to finally obtain a horizontal strip of excitation (250 mm X 150 mm) on the sample matching approximately the size of the non-collinear projection of the optical probe on the sample surface at all phase matching angles. The beam intensity provided by the SwissFEL was ranging from 300 mJ up to 800 mJ while the intensity at the sample was about 1.5 mJ. The phase gratings in the X-ray path were tilted in order to adjust their effective heights and then match the desired phase shift condition.

**Optical probe preparation** The optical probe laser was generated from a Ti:Sapphire laser delivering 35 fs pulses at 800 nm (10 mJ). A Barium borate (BBO) crystal was used to generate the second harmonic (400 nm) with an intensity of about 1.2 mJ. A bandpass filter (40 nm bandwidth) was used to remove the unwanted fundamental harmonic while a waveplate was used to control the intensity. Further filtering was passively done by several bandpass reflecting mirrors. The spot size at the sample was tuned to 190 x 150 mm2 (FWHM) by means of a lens. The arrival time of the probe laser was tuned by a delay stage upstream the sample and the final reflection to the sample was done by a D-shape mirror to accommodate for small phase matching angles. This D-shape mirror was mounted on a linear stage (see Fig. 2) in order to change the phase matching angle when different gratings were being used. The time overlap between the X-ray pump and the optical probe was readjusted for every phase-matching angle.

**Diffraction gratings fabrication** Phase gratings were made of polycrystalline CVD diamond supplied by Diamond Materials GmbH and used to excite transient gratings in the samples. The gratings were fabricated using a similar approach as reported by Makita et al [M1]. 10 $m$m-thick diamond membranes supported by a silicon frame were first cleaned in an $H_2SO_4:H_2O_2$ 2:1 solution at 120°C to remove any organic contamination. Then, the membranes were sputter-coated with a 10 nm thick Cr layer and subsequently spin-coated with a 1 mm thick negative tone resist FOX16 followed by baking at 100°C for 3 min on a hotplate. The resist was patterned in an electron beam lithography system (Raith EBPG 5000PlusES) using 100 kV accelerating voltage. After the exposure, the samples were developed for 8 min in Microposit 351:H20 1:3 solution at room temperature, then rinsed in DI water and isopropanol. The patterned resist gratings were hard baked at 300°C for 1 hour on a hotplate to increase etch selectivity between the resist and diamond. The unmasked Cr layer was removed in $Cl_2/O_2$ plasma revealing the underlying diamond for subsequent etching. Finally, the HSQ grating pattern was transferred into diamond by oxygen plasma etching in an Oxford PlasmaLab 100 machine with the following etching parameters: chamber pressure of 10 mTorr, 30 sccm O2 flow rate, powers of ICP and RF were 750W and 100W, respectively. After the etching, the remaining mask was stripped in 10% HF solution and the samples cleaned in Cr etchant and an $H_2SO_4:H_2O_2$ 2:1 solution, followed by rinsing in Di water and isopropanol.

**Detectors** The optical beams were measured by a Charged-Coupled Device (CCD) PCO-edge camera. A 2f-2f lens (200 mm half distance) was installed in front of the CCD imaging the sample at the chip. This allowed to reduce the background at the detector on the CCD camera. To control the angle of detection for different phase matching angles, the CCD camera was mounted on an heavy load $\delta - \gamma$ diffractometer at about 800 mm from the sample position and moved in the diffraction plane of the experiment. The $q$ angle was scanned in order to locate the diffracted XTG signal nearby the calculated value. A typical diffracted XTG spot recorded with the detector is displayed in Fig.4. The signal amplitude was obtained by integrating over the signal area while the weak background was subtracted using an area where the signal is absent. This weak background originated from the isotropic scattering of optical beam from the sample. Dark shots (unpumped signal) were recorded too and used for normalizing the data as well as to check that the sample was not printed during the measurements. Finally, an ultra-fast diode was positioned in transmission along the optical probe path with the purpose to measure the laser transmitted thorough the sample.

**References**

38 Mikako Makita et al. "Fabrication of diamond diffraction gratings for experiments with intense hard x-rays". In: Microelectronic Engineering 176 (2017), pp. 75–78

**Data availability** The raw data used in this study are available from the corresponding authors upon request.